\documentclass[twocolumn]{aastex63}
\accepted{April 23, 2022}
\received{}
\revised{}
\submitjournal{ApJ}

\usepackage{color}
\definecolor{orange}{rgb}{1,0.5,0} 
\definecolor{blue}{rgb}{0,0,0.6}  
\usepackage{apjfonts}
\usepackage{amsmath} 
\usepackage{lineno}

\usepackage{etoolbox}
\makeatletter
\patchcmd{\NAT@citex}
  {\@citea\NAT@hyper@{\NAT@nmfmt{\NAT@nm}\NAT@date}}
  {\@citea\NAT@nmfmt{\NAT@nm}\NAT@hyper@{\NAT@date}}
  {}
  {}
\patchcmd{\NAT@citex}
  {\@citea\NAT@hyper@{%
     \NAT@nmfmt{\NAT@nm}%
     \hyper@natlinkbreak{\NAT@aysep\NAT@spacechar}{\@citeb\@extra@b@citeb}%
     \NAT@date}}
  {\@citea\NAT@nmfmt{\NAT@nm}%
   \NAT@aysep\NAT@spacechar%
   \NAT@hyper@{\NAT@date}}
  {}
  {}
\patchcmd{\NAT@citex}
  {\@citea\NAT@hyper@{%
     \NAT@nmfmt{\NAT@nm}%
     \hyper@natlinkbreak{\NAT@spacechar\NAT@@open\if*#1*\else#1\NAT@spacechar\fi}%
       {\@citeb\@extra@b@citeb}%
     \NAT@date}}
  {\@citea\NAT@nmfmt{\NAT@nm}%
   \NAT@spacechar\NAT@@open\if*#1*\else#1\NAT@spacechar\fi%
   \NAT@hyper@{\NAT@date}}
  {}
  {}
\makeatother

\newcommand{\myemail}{takahiro@ipac.caltech.edu}
\newcommand{\simgt}{\,\rlap{\lower 3.5 pt \hbox{$\mathchar \sim$}} \raise
1pt \hbox {$>$}\,}
\newcommand{\simlt}{\,\rlap{\lower 3.5 pt \hbox{$\mathchar \sim$}} \raise
1pt \hbox {$<$}\,}
\newcommand{\Msun}{M_{\odot}}

\newcommand{\logm}{\log M_*/\Msun}

\newcommand{\kms}{{\rm km~s^{-1}}}
\newcommand{\oii}{[\textrm{O}~\textsc{ii}]}
\newcommand{\oiii}{[\textrm{O}~\textsc{iii}]}

\newcommand{\neiii}{[\textrm{Ne}~\textsc{iii}]}
\newcommand{\civ}{[\textrm{C}~\textsc{iv}]}
\newcommand{\ly}{${\rm Ly\alpha}$}
\newcommand{\ha}{${\rm H\alpha}$}
\newcommand{\hb}{${\rm H\beta}$}

\newcommand{\hst}{{HST}}

\newcommand{\Z}{{\rm [Z/H]}}

\newcommand{\ccc}{{3C~186}}

\newcommand{\sext}{{SExtractor}}

\newcommand{\fast}{{\ttfamily FAST}}

\begin{document}
\title{\Large
The Host Galaxy of the Recoiling Black Hole Candidate in 3C~186: An Old Major Merger Remnant at the Center of a z=1 Cluster
}

\newcommand{\affilA}{Space Telescope Science Institute, 3700 San Martin Drive, Baltimore, MD 21218, USA}
\newcommand{\affilB}{University of Florida, Department of Physics, 2001 Museum Rd., Gainesville, FL 32611, USA}
\newcommand{\affilC}{Department of Physics and Astronomy, University of Manitoba, Manitoba, Canada}
\newcommand{\affilD}{INAF - Osservatorio Astrofisico di Torino, Pino Torinese, Italy}
\newcommand{\affilE}{Dipartimento di Matematica e Fisica, Universit\`a degli Studi Roma Tre, via della Vasca Navale 84, 00146 Roma, Italy}
\newcommand{\affilH}{Dipartimento di Fisica e Astronomia ``Augusto Righi'', Alma Mater Studiorum Università di Bologna, Via Gobetti 93/2, I-40129 Bologna, Italy
}
\newcommand{\affilI}{INAF - Osservatorio  di  Astrofisica  e  Scienza  dello Spazio  di Bologna,  via  Gobetti  93/3,  I-40129,  Bologna, Italy}
\newcommand{\affilJ}{Leiden Observatory, University of Leiden, PO Box 9513, 2300 RA Leiden, The Netherlands}
\newcommand{\affilK}{Space Telescope Science Institute for the European Space Agency (ESA), ESA Office, 3700 San Martin Drive, Baltimore, MD 21218, USA}
\newcommand{\affilG}{The William H. Miller III Department of Physics and Astronomy, Johns Hopkins University, Baltimore, MD 21218, USA}
\newcommand{\affilL}{IPAC, California Institute of Technology, MC 314-6, 1200 E. California Boulevard, Pasadena, CA 91125, USA; \href{mailto:\myemail}{\myemail}}

\author{T.~Morishita}
\affiliation{\rm \affilA}
\affiliation{\rm \affilL}
\author{M.~Chiaberge}
\affiliation{\rm \affilK}
\affiliation{\rm \affilG}
\author{B.~Hilbert}
\affiliation{\rm \affilA}
\author{E.~Lambrides}
\affiliation{\rm \affilG}
\author{L.~Blecha}
\affiliation{\rm \affilB}
\author{S.~Baum}
\affiliation{\rm \affilC}
\author{S.~Bianchi}
\affiliation{\rm \affilE}
\author{A.~Capetti}
\affiliation{\rm \affilD}
\author{G.~Castignani}
\affiliation{\rm \affilH}
\affiliation{\rm \affilI}
\author{F.~D.~Macchetto}
\affiliation{\rm \affilA}
\author{G.~K.~Miley}
\affiliation{\rm \affilJ}
\author{C.~P.~O'Dea}
\affiliation{\rm \affilC}
\author{C.~A.~Norman}
\affiliation{\rm \affilG}
\affiliation{\rm \affilA}

\begin{abstract}
3C~186, a radio-loud quasar at $z=1.0685$, was previously reported to have both velocity and spatial offsets from its host galaxy, and has been considered as a promising candidate for a gravitational wave recoiling black hole triggered by a black hole merger. Another possible scenario is that \ccc\ is in an on-going galaxy merger, exhibiting a temporary displacement. In this study, we present analyses of new deep \hst/WFC3-IR and ACS images, aiming to characterize the host galaxy and test this alternative scenario. We carefully measure the light-weighted center of the host and reveal a significant spatial offset from the quasar core ($11.1\pm0.1$\,kpc). The direction of the confirmed offset aligns almost perpendicularly to the radio jet. We do not find evidence of a recent merger, such as a young starburst in disturbed outskirts, but only marginal light concentration in F160W at $\sim30$\,kpc. The host consists of matured ($\simgt200$\,Myr) stellar populations and one compact star-forming region. We compare with hydro-dynamical simulations and find that those observed features are consistently seen in late-stage merger remnants. Taken together, those pieces of evidence indicate that the system is not an on-going/young merger remnant, suggesting that the recoiling black hole scenario is still a plausible explanation for the puzzling nature of \ccc.
\end{abstract}

\section{Introduction}\label{sec:intro}
In the current paradigm of $\Lambda$-CDM cosmology, galaxy-galaxy mergers carry many important roles in galaxy evolution \citep{volonteri03,springel05}. Supermassive black hole (SMBH) merger is one of such expected as a result of major galaxy mergers. To explain tight relationships such as the $M$–$\sigma$ relation \citep{ferrarese00,gebhardt00}, an intimate relationship between galaxy growth and black hole growth during mergers seems to be required \citep[e.g.,][]{peng07}.

Black hole mergers are thought to happen in three phases: the two black holes are pulled towards the center of the gravitational potential of the merged galaxy by dynamical friction; they become a binary system by losing angular momentum via gravitational slingshot interaction with stars that have appropriate angular momentum in the region of the parameter space \citep[the so-called loss cone;][]{frank76,begelman80} and gas-driven inspiral \citep[e.g.,][]{escala05,dotti07}; in the final phase, the bound pair may lose the remaining angular momentum via the emission of gravitational waves (GWs), and eventually, the two black holes coalesce. In most cases, these GWs are emitted anisotropically. Depending on both the relative orientation of the spins of the merging black holes and their mass ratio, the merged black hole may receive a recoil kick \citep{centrella10,blecha11,komossa12} with velocities as high as $\sim5000$\,km\,s$^{-1}$ \citep[e.g.,][]{campanelli07,tichy07,lousto11}. As a result, the merged black hole may show displacement from the center of the host galaxy both in position and velocity.

However, the details of the mechanisms that bring the two black holes to the distance at which GW emission becomes substantial are still poorly understood. This is called the final-parsec problem \citep[e.g.,][]{milosavljevic03}. For example, if the loss cone is not replenished quickly enough for the SMBH pair to get sufficiently close and lose energy via GWs, the pair may stall and never merge; those may be observed as dual quasars or pairs of less active AGN \citep[e.g.,][]{komossa03,hennawi06,goulding19,silverman20,oneill21}. If this is often the case and prevents SMBH pairs to merge, our current understanding of the mechanisms for black hole growth may need significant revisions. It is thus important to find, either direct (GWs) or indirect (recoiling black holes), evidence of SMBH mergers. While we still need to wait for next generation space-based GW interferometers and pulsar timing arrays for the direct confirmation at the black hole mass range of $M_{\rm BH}\simgt10^8M_\odot$ \citep[see, e.g.,][]{moore15}, identifying recoiling black holes helps us to advance our understanding of such extreme events.

As of today, there are several candidates of recoiling black holes, with velocity \citep{komossa08,steinhardt12,comerford14,pesce18}, spatial \citep{batcheldor10,koss14,lena14,barrows16,skipper18}, or both offsets \citep{pesce21,hogg21}. In particular, high-velocity offsets ($>1000$\,km\,s$^{-1}$) are expected to be rare, but they are more likely to be observed in combination with large spatial offsets \citep[e.g.,][]{blecha16} and thus can be used as a proxy for ideal followup targets.

For example, CID-42, a radio-quiet AGN at $z=0.36$, shows a velocity offset between the narrow and broad components of the \hb\ emission line of $\sim1300$\,km/s, with two nuclei in HST images displaced by $\sim2.5$\,kpc in projected distance \citep{civano10,civano12,blecha13,novak15}. The presence of a second, obscured radio quiet AGN, however, cannot be excluded \citep[but see also][]{kim17}. Another intriguing case is found in A2261-BCG, the brightest cluster galaxy (BCG) in Abell~2261 at $z=0.2248$. This galaxy was found to have an unusually flat core profile, with an offset of $\sim0.7$\,kpc from the photocenter of the surrounding envelope \citep{postman12b}. The followup study presented spectroscopic measurements of three of the four knots identified in the central region, to test their hypothesis that the massive black hole was ejected from the core and harbour within one of those knots \citep{burke-spolaor17}. Their observations could not identify the precise location of the expected SMBH ($\sim10^{10}\,M_\odot$), leaving the conclusion still pending. 

\ccc\ is another promising candidate, but by far at higher redshift than others. The system is a radio-loud quasar \citep[$\sim10^5$\,yrs;][]{murgia99,siemiginowska05} of the compact-steep spectrum class \citep{fanti85,odea98,odea21},  located in the center of a cluster of galaxies at $z=1.0685$, with a SMBH of $\sim3$-$6\times10^9\,M_\odot$ \citep{kuraszkiewicz02,chiaberge17}.

\ccc\ was recently observed as one of 22 3C~radio galaxies/quasars in a snapshot program of HST \citep{hilbert16}. \citet[][]{chiaberge17} reported its spatial offset by $1.3\pm0.1$\,arcsec, or $\sim 11$\,kpc, with respect to the center of the host galaxy seen in their medium-deep image of WFC3-IR F140W. In the same study, \ccc\ was also revealed to have a significant velocity offset ($\sim2140$\,km/s) of broad lines associated with the quasar with respect to the systemic redshift of the host measured by two narrow lines, \oii\ and \neiii, suggestive of a binary black hole merger of a mass ratio $m_1/m_2 > 0.25$ \citep{lousto17}. In addition, the followup observations with Keck/OSIRIS integral field unit (IFU) revealed a tentative but significant velocity shift ($>1800$\,km/s) in the broad component of \hb\ line from narrow lines \citep[][see also Sec.~\ref{ssec:spec}]{chiaberge18}.

While all previous results support the interpretation of \ccc\ as a GW recoiling black hole, there are two other possible scenarios that may explain the observed properties; one is that we are seeing an on-going galaxy merger, where its photocenter is temporarily displaced due to disturbances in the host morphology; the other scenario is that the quasar is associated with an under-massive galaxy that is superimposed to another brighter galaxy. 

With the depth of the previous F140W image, those scenarios were not completely excluded, as tidal features and recent star formation in the outskirts can easily be missed \citep[e.g.,][]{koss18}. Deeper imaging with \hst\ is thus important to unveil any evidence of on-going merger and constrain the current stage of the system. 

The aim of this paper is to test these alternative hypotheses by analysing deeper IR and optical images newly taken in the HST Cycle~25. In particular, high photometric sensitivity of \hst\ at $\sim1$\,kpc resolution provides details of the underlining stellar population of the host galaxy and allows us to further constrain the time scale since the last major merger and associated star formation. 

Throughout, we adopt the AB magnitude system \citep{oke83,fukugita96}, cosmological parameters of $\Omega_m=0.3$, $\Omega_\Lambda=0.7$, $H_0=70\,\kms\, {\rm Mpc}^{-1}$, and the \citet{salpeter55} initial mass function. We refer to magnitude for the HST filters used in this study, F606W, F110W, and F160W as $V_{606}$, $J_{110}$, and $H_{160}$ respectively.

\begin{figure*}
\centering
	\includegraphics[width=0.9\textwidth]{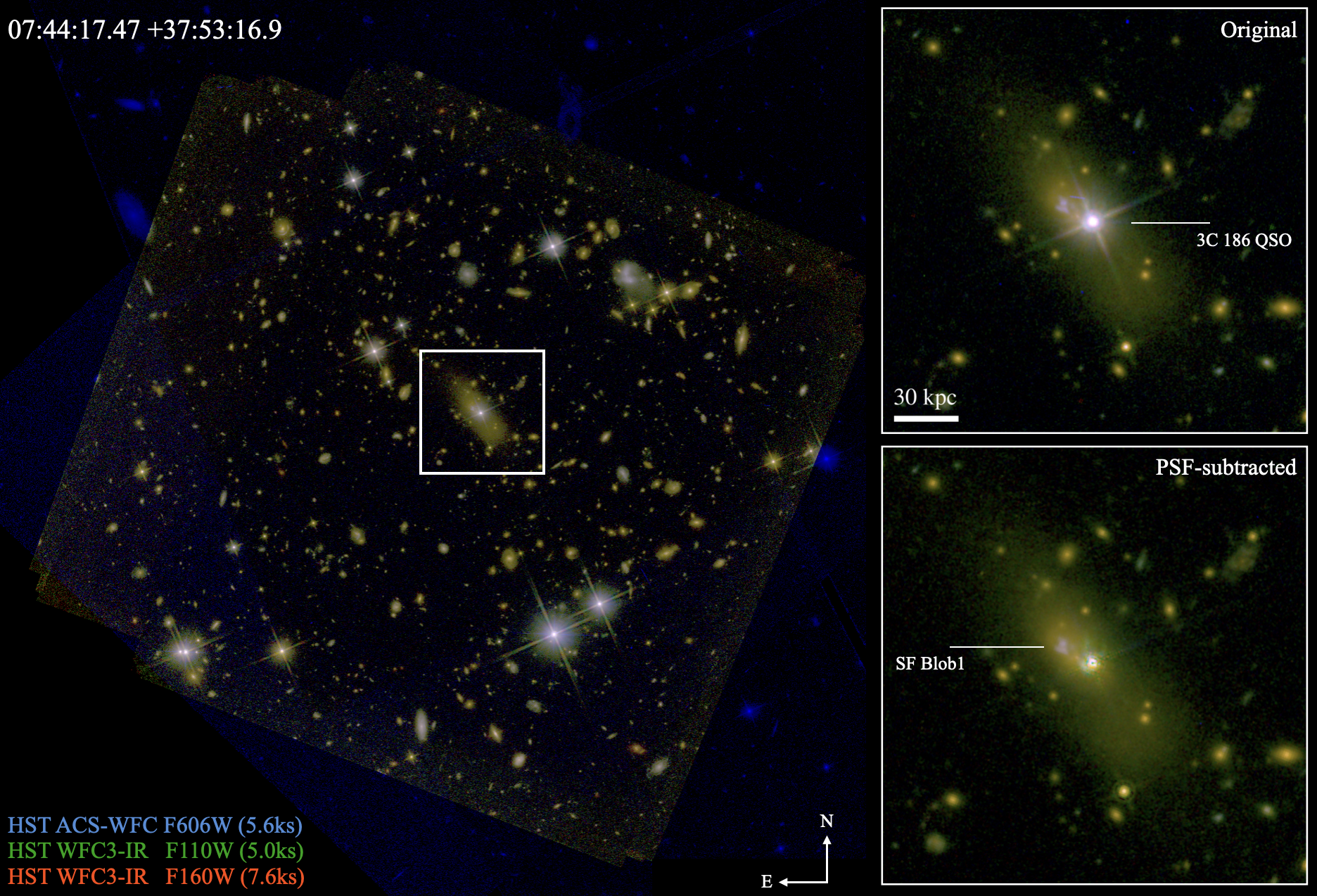}
	\caption{Pseudo color image of the \ccc\ field (F606W for blue, F110W green, F160W red). A zoomed image of \ccc\ ($\sim27\times27$\,arcsec$^2$) is shown in the two right panels, in the original (top) and PSF-subtracted (bottom) images. The compact star-forming blob, SF Blob1 (Sec.~\ref{ssec:bb}) is indicated in the right-bottom panel. The images have been rotated so that north is up and east to the left.
	}
\label{fig:mosaic}
\end{figure*}

\begin{deluxetable}{lcc}
\renewcommand{\arraystretch}{0.99}
\tabletypesize{\small}
\tabcolsep=7pt
\tablecolumns{3}
\tablewidth{0pt}
\tablecaption{Summary of HST observations in the \ccc\ field.}
\tablehead{
	\colhead{Instrument} &
	\colhead{Filter} &
	\colhead{Total Exposure Time}
\vspace{-0.3cm}\\
	\colhead{} & 
	\colhead{} & 
	\colhead{sec} 
}
\startdata
ACS-WFC & F606W & 5567\\
WFC3-IR & F110W & 5042\\
WFC3-IR & F160W & 7563
\enddata
\end{deluxetable}\label{tab:obs}

\section{Data}\label{sec:data}

\subsection{\hst\ Observations and Data Reduction}\label{ssec:hst}
\ccc\ was observed in a GO program (PID 15254; PI. M. Chiaberge) during the \hst\ Cycle 25, with WFC3-IR and ACS. Two filters, F110W and F160W, are used in the WFC3 observations and one filter, F606W, in the ACS observations (Table~\ref{tab:obs}). The observing position angle was carefully designed during the phase~II process, so to minimize possible contamination on the host galaxy by stellar spikes from the quasar and surrounding objects. To avoid saturation of a few pixels in the quasar core, we started with one short exposure ($\sim30$\,sec) in each of the 5~orbit ACS observations. Then, the following exposures were taken at the same location and rotation angle of the first dither point in each orbit, to ensure an identical PSF profile over the entire exposures.

To reduce imaging data, we follow the procedure presented in \citet{hilbert16}. For ACS data, we begin with files in the {\ttfamily flc} format. These images are corrected for charge transfer efficiency (CTE) effects through the CTE correction algorithm \citep{anderson10}. For WFC3-IR data, we begin files in the {\ttfamily flt} format, as the multiple non-destructive detector readouts within each observation allowed for easy cosmic ray identification and removal within the calwf3 data reduction pipeline. However, pixels subjected to high flux levels in preceding observations are affected by persistence and need to be corrected. To remove any persistence signal present in our data, we begin by retrieving the persistence masks and persistence-corrected flt files of our observations from MAST. These files have had the persistence signal modeled and subtracted from them following the model described in Section 8.3 of version 4.0 of the WFC3 Data Handbook, and are therefore different from the standard flt files available in the archive. Before proceeding with the data reduction, we also manually mask the satellite trail present in one of our images.

Since our focus in this study is the host galaxy of \ccc, we attempt to subtract the light from the bright quasar nucleus in each distortion uncorrected image (those in flt/flc formats) before the geometric correction and final stacking steps (see Sec.~\ref{ssec:psf}).

The next step in our data reduction is to remove the geometric distortion from all of the PSF-subtracted flc/flt files and to combine images into a final image for each filter. We use Astrodrizzle to accomplish both of these steps. 

For each filter, we combine the PSF-subtracted images into a final image with a pixel scale of 0.045"/pixel. We find a value of "5.5 5.0" for the \emph{driz\_cr\_snr} parameter results in good cosmic ray rejection during the stacking process. We also calculate an optimal \emph{final\_pixfrac} value of 0.75 for our data and final pixel scale.

We then use Tweakreg (also part of the Drizzlepac software package) to align this final image to the same world coordinate system (WCS) present in the corresponding ACS drizzled image. At this point, it is possible to overlay the ACS and WFC3 images for a particular object and compare the morphology and brightness in the two observation bands.

The resulting final image is shown in Fig.~\ref{fig:mosaic}. The PSF-subtracted image reveals a clear spatial offset between the quasar and the host galaxy (Sec.~\ref{ssec:lwc}). Our deep images also reveal faint outskirts of the host, out to $\sim60$\,kpc.

\subsection{PSF Subtraction}\label{ssec:psf}
In order to subtract the PSF component without over-subtracting the host light component, we fit the observed light profile with PSF+single S\'ersic profile using GALFIT \citep{pengGALFIT}. In an effort to achieve the best model fits, we experiment with the GALFIT inputs. We create a custom bad pixel file for each exposure, where we remove all nearby astrophysical sources as well as bad pixels and cosmic rays. By supplying this file to GALFIT, we reduce the chances that the final fit is contaminated by nearby sources.

The PSFs used for our PSF subtractions are created using TinyTim software \citep{krist11}. We use TinyTim PSFs due to a lack of unsaturated, high-SNR, isolated PSFs in our data to use as models. In order to create realistic PSFs, we used characteristics from our observations when creating the model PSFs. We begin by finding the mean secondary mirror focus offset for the date and time of our observation. This information is collected from the HST focus model website\footnote{\url{http://www.stsci.edu/hst/observatory/focus/FocusModel}}.

Other information used as input to TinyTim includes the \hst\ instrument and filter, the target's pixel location on the detector, the amount of jitter (0 for this work), and the size in arcseconds of the output model PSF (7\,arcseconds). Finally, we also provide a quasar spectrum, in order for Tiny Tim to determine wavelength-dependent PSF characteristics. For simplicity, we used the composite quasar near-IR spectrum from \citet{glikman07} redshifted to $z\sim1$. With this information in hand for each exposure, we run all three stages of TinyTim. The final output is one geometrically distorted model PSF corresponding to each of flc/flt images.

We also fit the data using a PSF at the nominal pixel scale, as well as a supersampled PSF. We find that the fit with the supersampled PSF tends to over-subtract the source flux, and therefore keep to the nominal pixel scale for our final fittings. We also fit the data with TinyTim PSFs from a range of focus values, in order to examine any differences. The best modeling and subtraction came from values that match the focus values indicated by the focus model website. Finally, as the data being fit are in units of electrons per second, rather than the electrons that GALFIT assumes, we also experiment with fits where we provide the ERR array of the flt/flc file to GALFIT as a sigma image. We find in this case that the resulting fits again tend to over-subtract the source's flux. 
With our final set of GALFIT parameters and inputs, we obtain fits with reduced-$\chi^2$ between 0.75 and 1.5. The best-fit PSF component is then subtracted from each of the images before they are processed for geometric correction and final stacking with Astrodrizzle.

\begin{figure*}
\centering
	\includegraphics[width=0.33\textwidth]{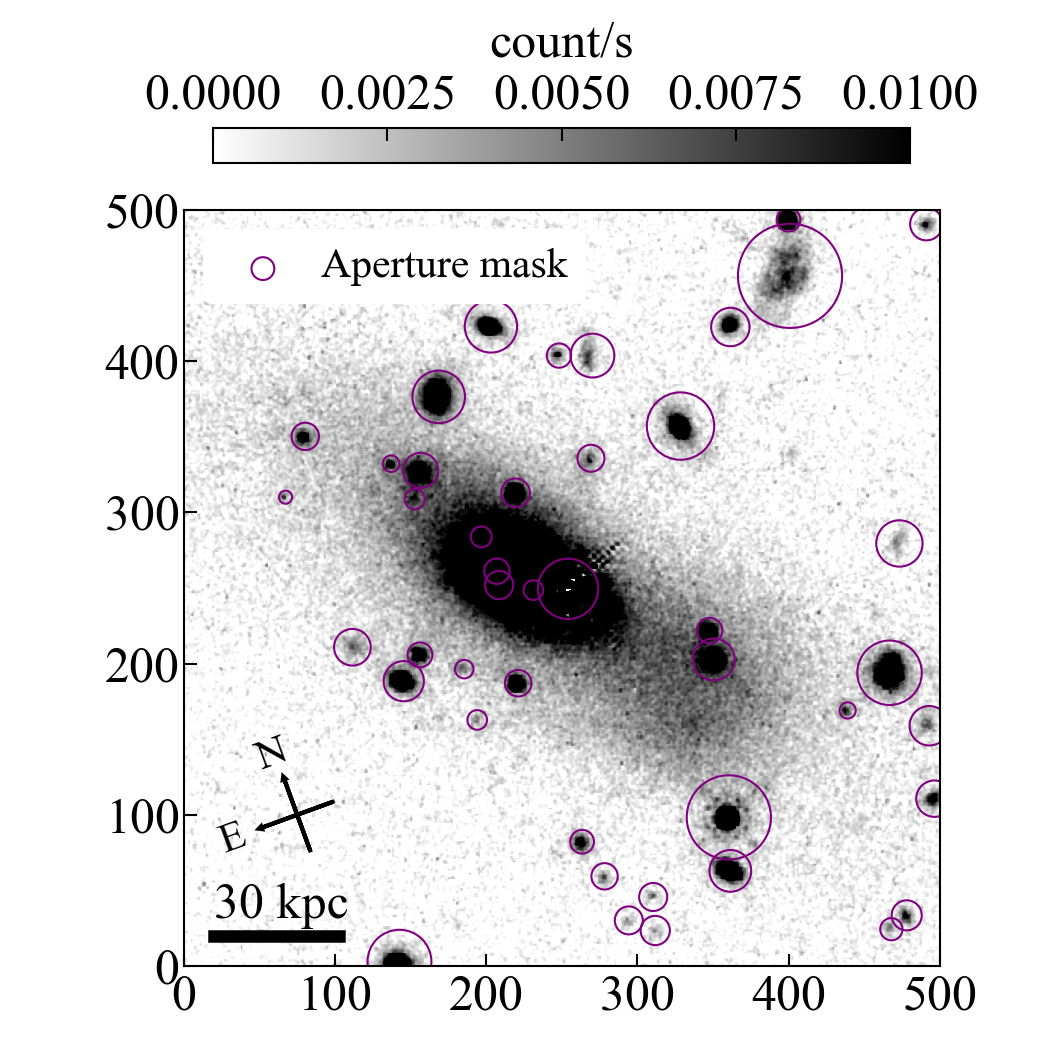}
	\includegraphics[width=0.33\textwidth]{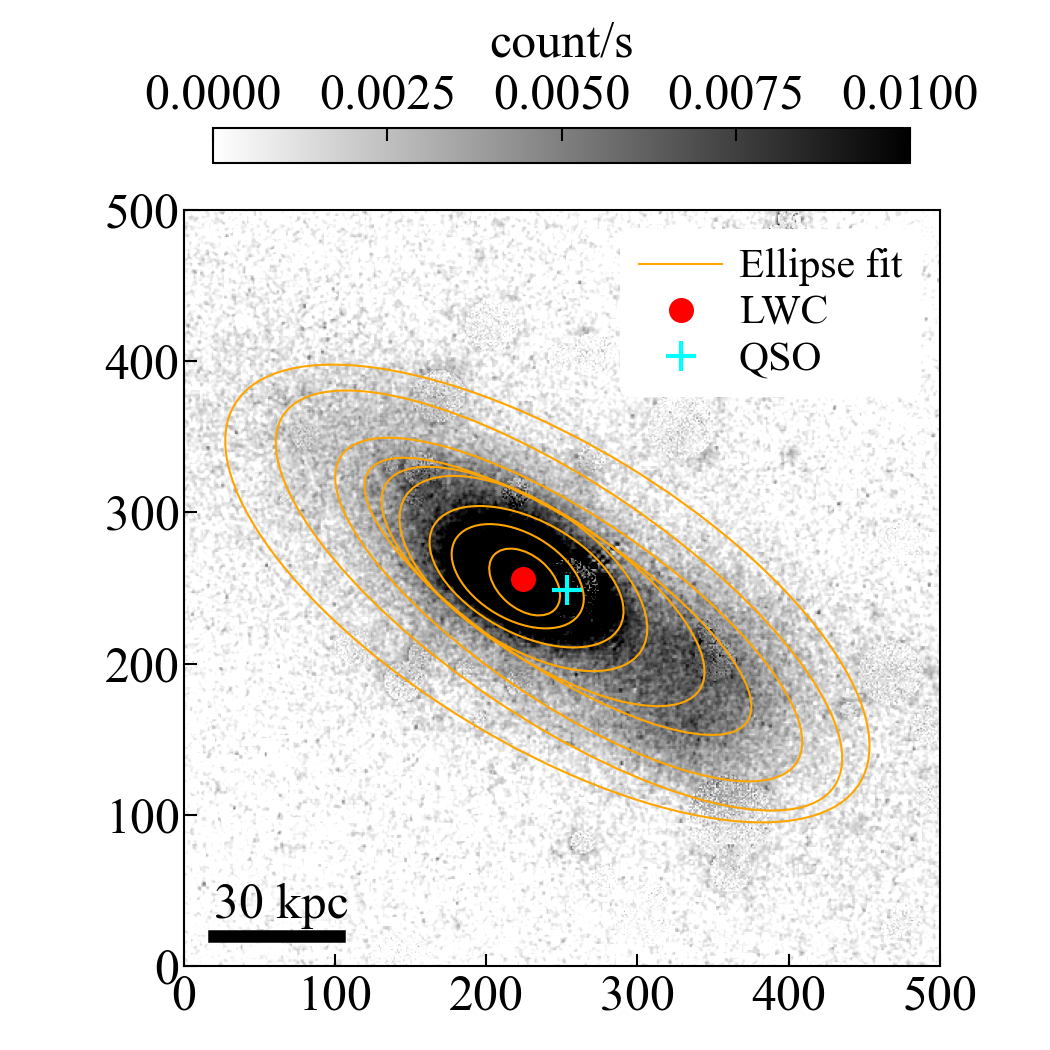}
	\includegraphics[width=0.33\textwidth]{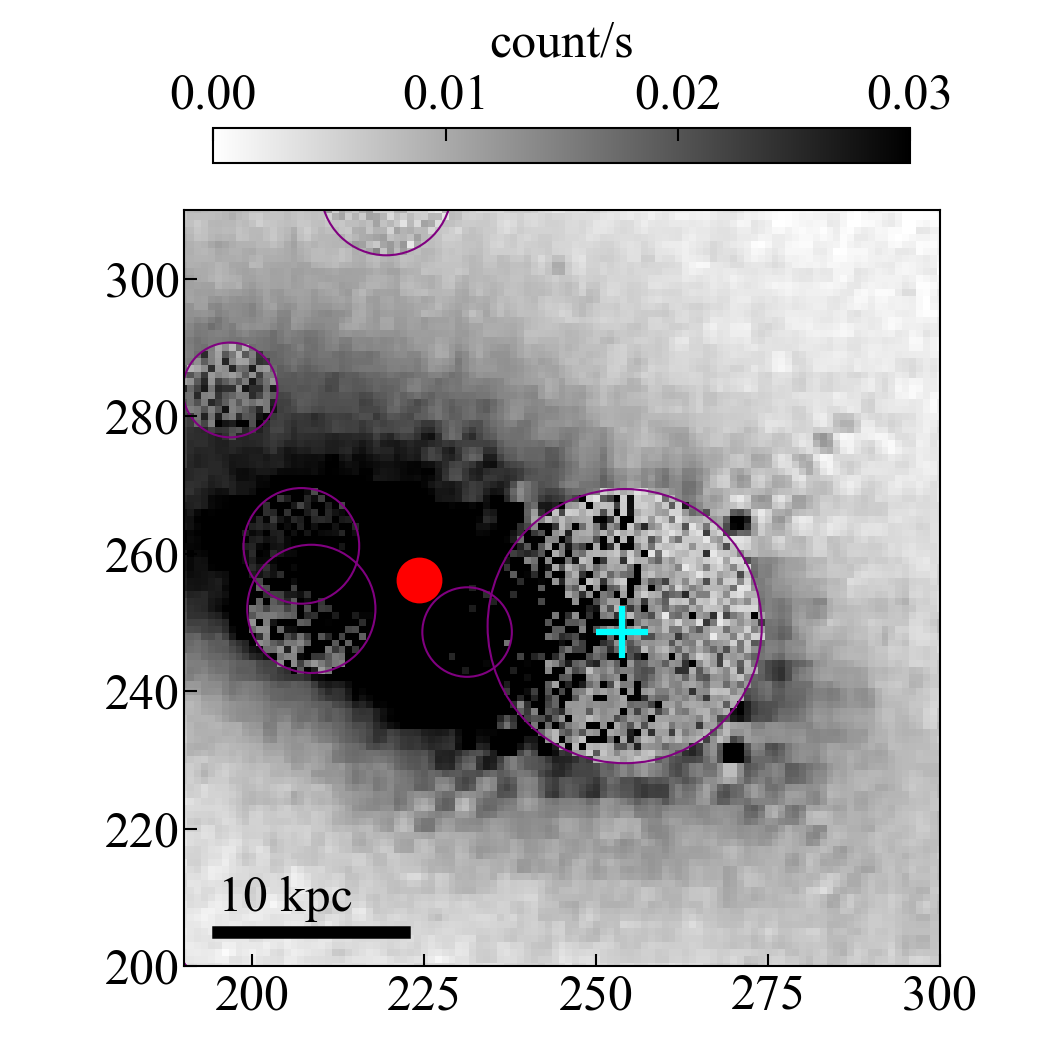}
	\caption{
	{\it Left}: PSF-subtracted image of \ccc\ in F160W, in a $500\times500$\,pixel postage stamp. Regions that are masked and patched are indicated with purple apertures.
	{\it Middle}: One of the patched images, where all masked regions are filled by random values taken from nearby pixels (Sec.~\ref{ssec:lwc}). The median light-weighted center (red point), calculated over 100 such realizations, is clearly off from the flux peak position of the quasar (cyan cross). Fitted ellipses are shown (orange lines). Light concentration is seen in the South-West direction from the quasar.
	{\it Right}: Zoom-in image of the middle panel with different color stretch.
	}
\label{fig:lwc}
\end{figure*}

\section{Analysis and Results}

\subsection{Light-weighted center of the host galaxy}\label{ssec:lwc}
We estimate the center of mass of the host galaxy by using F160W, which is the deepest among our images here. It is noted that the light distribution in the F160W image is considered to trace the stellar component of the host well, as at the redshift of \ccc\ the filter corresponds to rest-frame $\sim0.8\,\mu$m and is clean from strong emission lines (see also Sec.~\ref{ssec:sed}).

While the extended light from the host galaxy is clearly visible in the PSF-subtracted image (Fig.~\ref{fig:mosaic}), it is still challenging to accurately estimate its light-weighted center (LWC) due to the fact that fore/background objects are superposed on the host. Especially, since \ccc\ is in a crowded field, the latter remains as a critical issue because flux contribution from other sources can easily affect the estimate.

To eliminate such flux contamination, we first manually mask regions that are likely to belong to surrounding objects ({including the compact star-forming blob discussed in Sec.~\ref{ssec:bb}}) and residuals at the position of quasar, by placing circular apertures of arbitrary sizes in the PSF-subtracted F160W image. By doing so, we ensure to mask all possible contaminants, whereas automated software (e.g., \sext; \citealt{bertin96}) may not necessarily recognize all objects in crowded regions or mistakenly mask the part of the host. The image with those aperture masks is shown in the left panel of Fig.~\ref{fig:lwc}. {To mask the residuals at the location of the quasar, we place an aperture with radius of $r=0.\!''9$.}

The next step is then to fill those masked regions. This process is necessary, because leaving those masked pixels could result in a biased estimate of LWC depending on the choice of position and size used of each circular mask. For each pixel of the masked regions, we use a Metropolis-Hastings-like algorithm to randomly select one of the neighboring pixels that are not masked, to fill in the target pixel. This filling process is as follows;
\begin{itemize}
\item[] 1.~For the pixel of interest in a masked region, assign a radius ($r_{rand}$) randomly taken from a range of $[0, r_{max}]$, where we here set $r_{max}=50$\,pixel.
\item[] 2.~Compare the probability of $r_{rand}$, assigned by $p(r)=\exp{(-r / r_{max})}$, with a random float number, $p_r$, drawn from a range of [0,1].
\item[] 3.~If $p(r_{rand})>p_r$, then select a random pixel from unmasked regions at $r_{rand} < r <r+\Delta r$, where we set $\Delta r=2$\,pixel; otherwise return to 1 and repeat the process until it passes the criteria.
\item[] 4.~Repeat step 1-3 for all masked pixels.
\item[] 5.~Repeat step 1-4 for $n_{iter}$ times, to get $n_{iter}$ images with the masked regions filled respectively.
\end{itemize}

We here set $n_{iter}=100$. By adopting an exponential function for $p(r)$, the filling flux is likely to be chosen from one of the nearby pixels, making the reproduced light profile contiguous, which is reasonable given the primary purpose here. One of the realizations is shown in Fig.~\ref{fig:lwc} as an example. While several pixels can occasionally have a distinct value from the average value of surrounding pixels, contributions from such pixels to the determination of LWC are smoothed out after repeating $n_{iter}$ times of realization.

At each realization of $n_{iter}$, we calculate the LWC by the following equations;
\begin{equation}
	\begin{aligned}
            \overline{x} = {{\sum x f_{\rm 160}(x,y) / e_{\rm 160}(x,y)^2}\over{\sum f_{\rm 160}(x,y) / e_{\rm 160}(x,y)^2}},\\
            \overline{y} = {{\sum y f_{\rm 160}(x,y) / e_{\rm 160}(x,y)^2}\over{\sum f_{\rm 160}(x,y) / e_{\rm 160}(x,y)^2}}
	\end{aligned}
\end{equation}
where $f_{\rm 160}(x,y)$ and $e_{\rm 160}(x,y)$ represent flux value and its associated RMS value at the position of [x,y], and the sum goes over pixels with $f_{\rm 160}/e_{\rm 160}>{\rm SNR_{limit}}$ within $r_{\rm lim}<250$\,pixel ($\sim91$\,kpc) from the target. {We here set ${\rm SNR_{limit}}=0.7$, which is equivalent to the detection threshold ({\tt DETECT\_THRESH} of \sext) used for source detection in the F160W image. We empirically find that the value in general provides reasonable detection out to faint outskirt of sources without being contaminated by background pixels.}
 
\begin{figure}
\centering
	\includegraphics[width=0.38\textwidth]{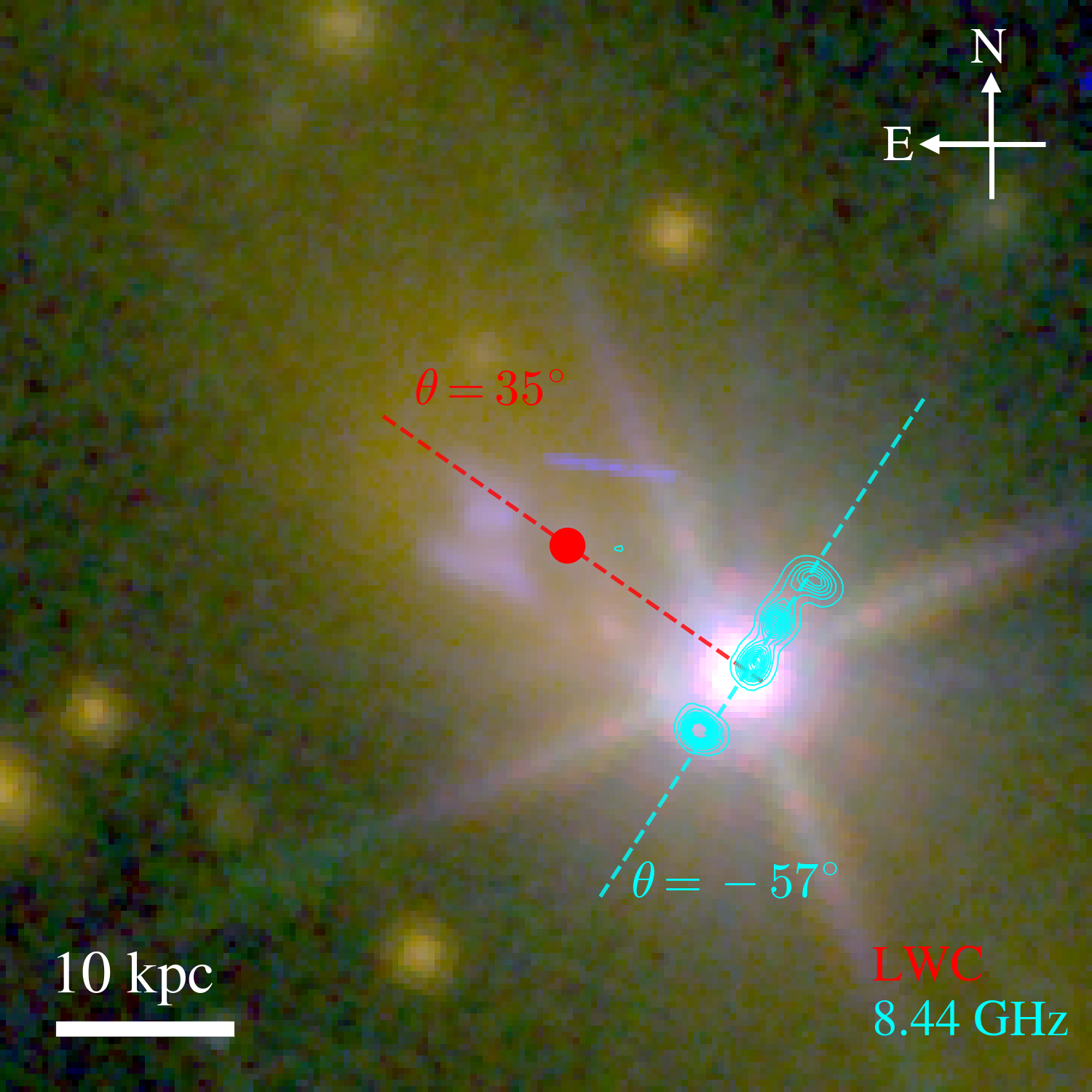}
	\caption{Central region of \ccc. The RGB image is in the same configuration as in Fig.~\ref{fig:mosaic}. The position of the light-weighted center is shown (red circle). The 8.44\,GHz radio continuum emission, retrieved from the NRAO VLA archive (project ID AA0129; cyan contour), is overlaid. Two dashed lines are shown to highlight the inferred axes of the offset (red) and the jet (cyan), respectively. The image has been rotated so that north is up and east to the left.
	}
\label{fig:radio}
\end{figure}

We calculate the median LWC over $n_{iter}$ realizations and find 
$[x,y]=[-29.42_{-0.08}^{+0.07}, 7.44_{-0.05}^{+0.06}]$ (in pixel) 
with respect to the position of \ccc\ in the image of Fig.~\ref{fig:lwc}, or 7:44:17.586,$+$37:53:18.165 in the sky coordinate. The 16th/84th percentiles are associated as uncertainties. The result reveals a significant offset $1.37\pm0.01$\,arcsec, or $11.1\pm0.1$\,kpc, in the projected distance, which is consistent with the estimate with a shallower F140W image by \citet{chiaberge17}. The angle from the LWC toward the quasar core, measured from the west to the clockwise direction, is $\sim+35^{\circ}$, which lies almost perpendicularly to the direction of the radio jet (Fig.~\ref{fig:radio}). We will discuss this in Sec.~\ref{ssec:dyn}.

The choice of $r_{\rm lim}$ and the limiting SNR has only a minor effect on the estimate as long as ${\rm SNR_{limit}}$ is set to $>0.7$; having lower limiting SNR would significantly increase the estimated error due to noisy pixels across the entire image.
On the other hand, the inferred offset becomes even larger when a higher value for ${\rm SNR_{limit}}$ is chosen. As in the middle panel of Fig.~\ref{fig:lwc}, ellipse fit to the F160W light profile shows a shift of the centroid toward the North-East direction when only the central part of \ccc\ is used. This is due to the light concentration in the outskirts. We discuss the light concentration in relation to merger-induced morphological disturbances, or lack thereof, in Sec.~\ref{sec:dis}.

Despite the depth that our new images reach, we might still be missing very low-surface brightness components ($>25$\,mag\,arcsec$^{-2}$). However, we argue that such light components only marginally affect the determined photocenter. Our images here already reveal a substantial amount of stellar mass\footnote{Such an amount of mass may be found in the diffuse stellar components but only in the most massive clusters \citep[e.g.,][]{morishita17}.} ($\logm\sim11.4$; see Sec.~\ref{ssec:sed}). The amount is $\times10$ more than what is found in the diffuse outskirts of low-$z$ elliptical galaxies \citep[e.g.,][]{huang18}.

\subsection{Stellar populations of the host galaxy}\label{ssec:sed}
We aim to investigate the underlying stellar population in the host. The spatial distribution of stellar age enables us to estimate the timeline since the last merger and associated star formation activity. 

\begin{figure}
\centering
	\includegraphics[width=0.48\textwidth]{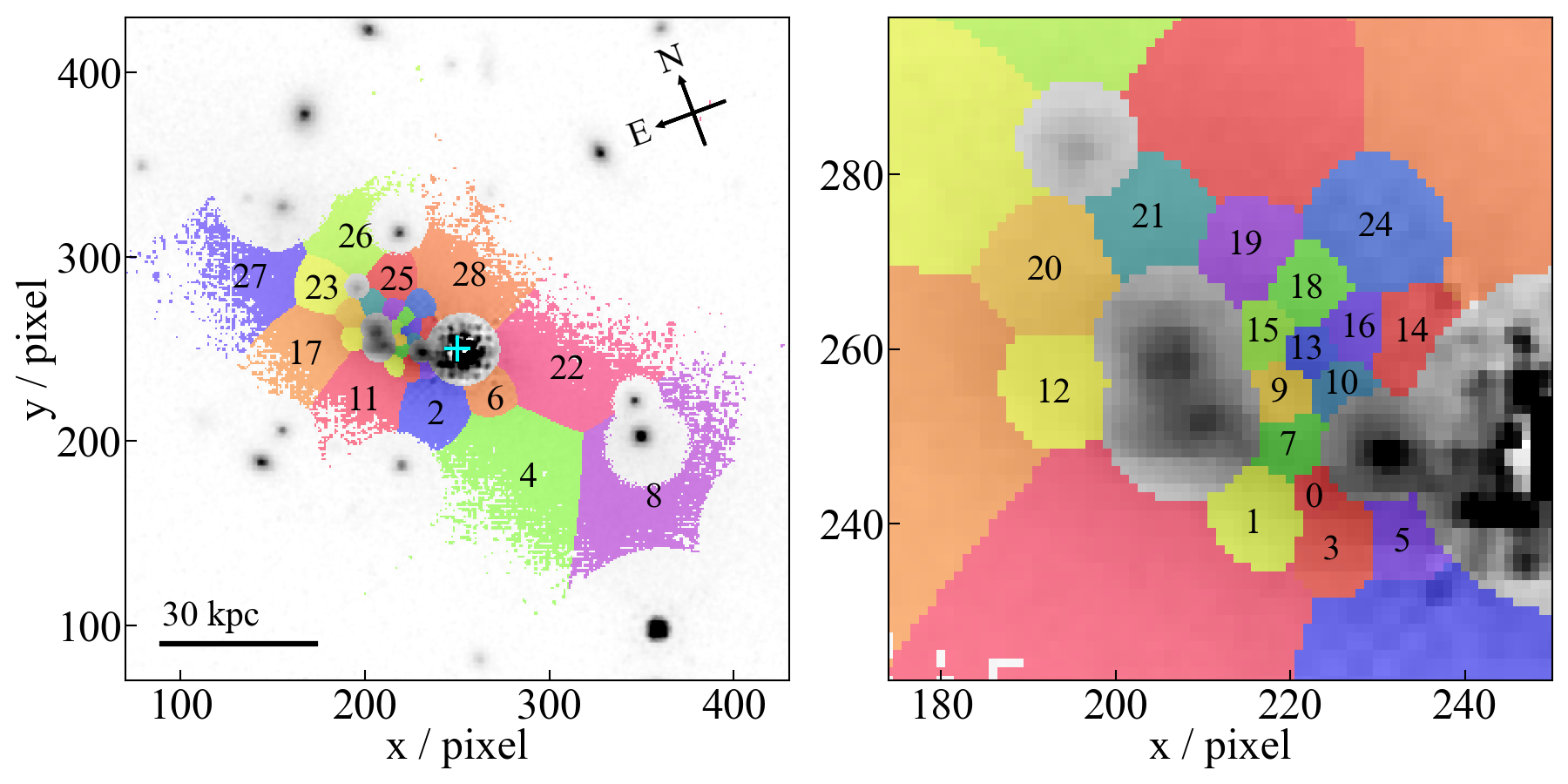}
	\caption{
	{\it Left}: Segmentation map, where each tessellated region is color-coded differently, overlaid on the F160W image. 
	IDs of each segment used in the main text are shown (also, Table~\ref{tab:sed}). The position of \ccc\ is marked by a cyan cross symbol.
	{\it Right}: Central region of the same Voronoi segmentation map. The region of the blue blob at [x,y] = [210,260] (Sec.~\ref{ssec:bb}) is excluded from the host population analysis.
	}
\label{fig:vor}
\end{figure}

To get sufficient SNR for reliable estimates out to faint outskirts of the host, we define sub-regions of the host by using a Voronoi tessellation method \citep{cappellari03}. We set the minimum SNR to $15$, so that every tessellated region has SNR greater than this value. As a result, at $\sim6$\,arcsec ($\sim48$\,kpc) from the system LWC we reach down to $\sim25$\,mag\,arcsec$^{-2}$ in F160W. We set the boundary of the host by using the segmentation mask generated by \sext\ with the detection threshold of $0.7$, which is consistent with the threshold used for the LWC calculation above. Note that in this analysis, as it is not necessary for our purposes, we do not fill the masked regions defined in Sec.~\ref{ssec:lwc} and leave those empty. We show the defined mask and radial distribution of tessellated pixels in Fig.~\ref{fig:vor}.

The boundary for tessellation defined in the F160W image are then applied to both F110W and F606W images, to collect flux from the same sub-regions. Each of the F110W and F606W images is convolved to the PSF size of the F160W image beforehand. The convolution kernels were generated by providing median-stacked PSFs to a python software, {\ttfamily pypher} \citep{boucaud16}, in the same way as in \citet{morishita21}. Fluxes of each sub-region are summed and compiled into a catalog for a spectral energy distribution (SED) analysis in the following section.

\begin{figure}
\centering
	\includegraphics[width=0.45\textwidth]{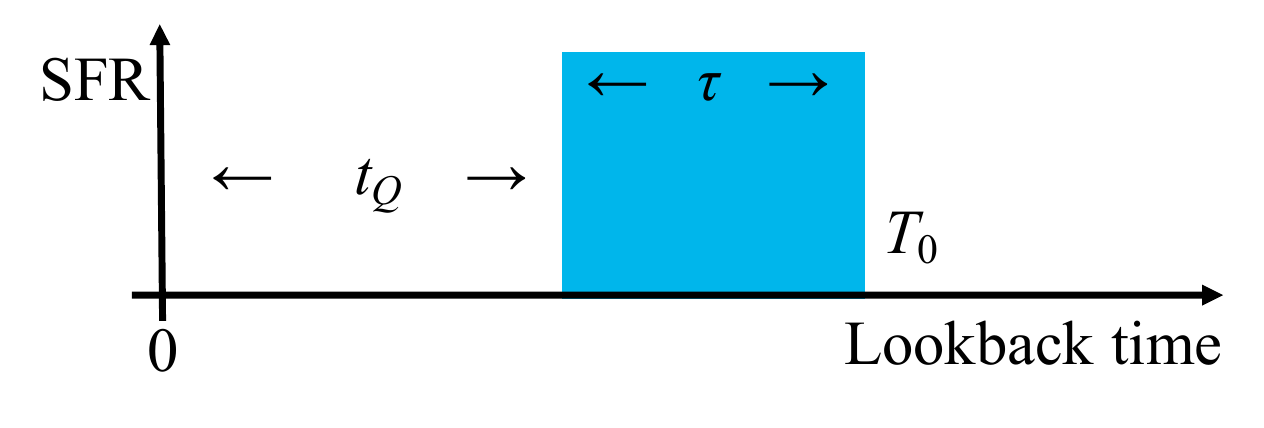}
	\includegraphics[width=0.47\textwidth]{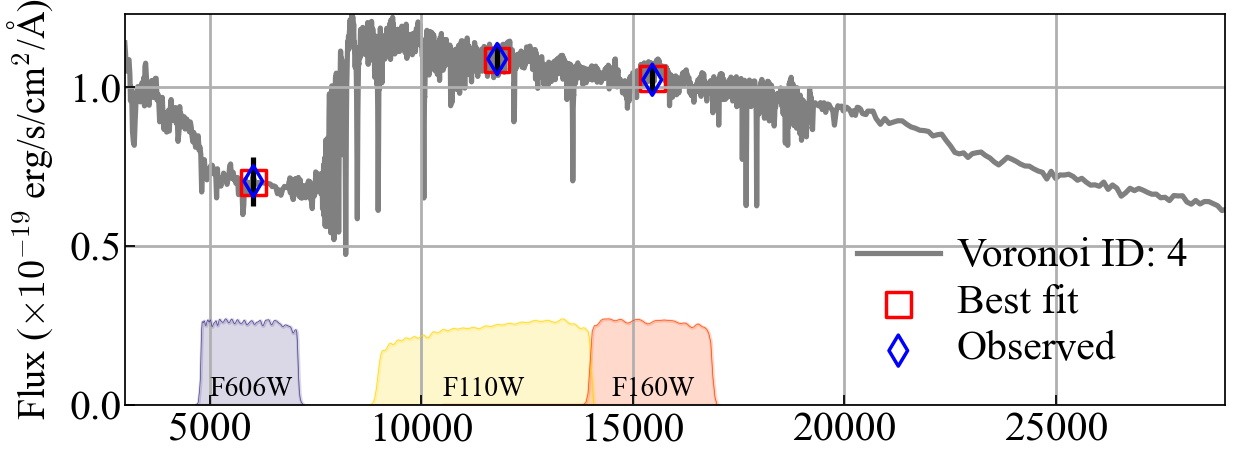}
	\includegraphics[width=0.47\textwidth]{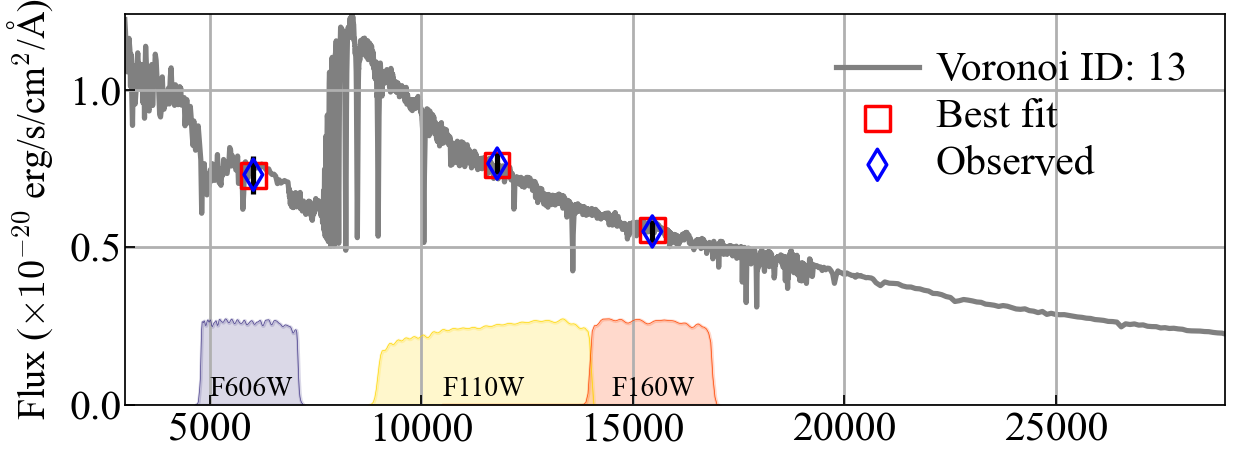}
	\includegraphics[width=0.47\textwidth]{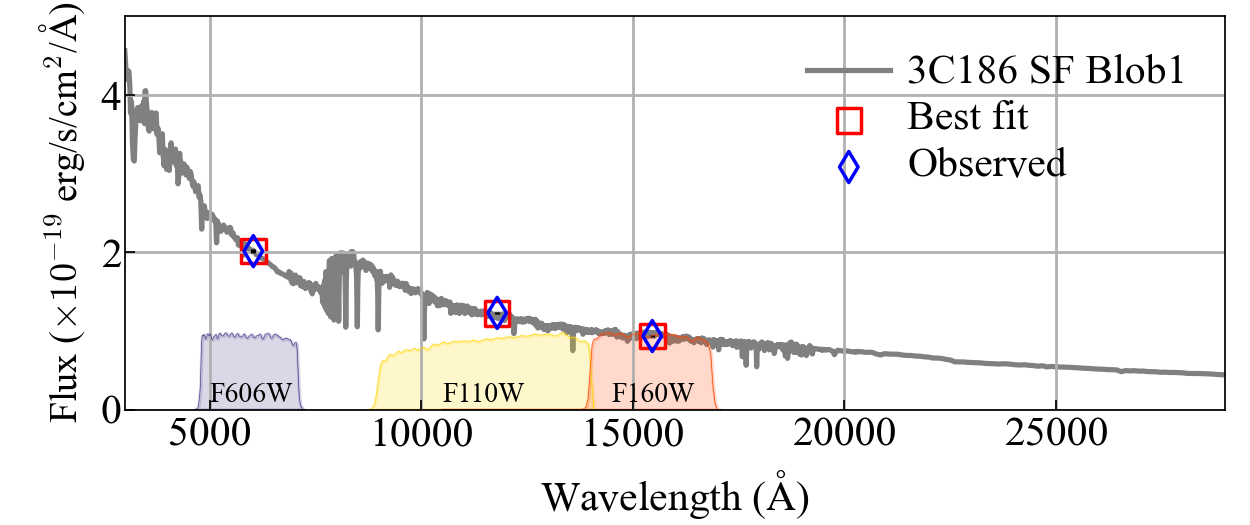}
	\caption{
	{\it Top}: Schematic of our star formation history model, where star formation initiates $T_0$ ago, constantly lasts for over time of $\tau$, and then terminates at $t_Q$ ago.
	{\it Bottom}:
	Examples of SED fitting results of three regions: Voronoi ID 4 (top), Voronoi ID 13 (middle), and SFB1 (bottom). The best-fit SEDs (solid lines) and convolved data points (red squares) are shown. Despite the number of observed data points (3; blue diamonds), the data constrain the model by capturing the Balmer break ($V_{\rm 606}-J_{\rm 110}$; sensitive to stellar age) and optical slope ($J_{\rm 110}-H_{\rm 160}$; sensitive to dust). It is noted that at the redshift of \ccc, F160W is clean from strong emission lines such as \ha, \hb, and \oiii, making it a good tracer of the stellar component.
	}
\label{fig:sed}
\end{figure}

The flux catalog collected from the tessellated images is provided to an SED fitting software, \fast\ \citep{kriek09}. Due to the number of broadband filters available here (3), we are limited to a small number of fitting parameters. Since our primary focus here is to estimate the epoch of the last significant star formation (which could potentially be associated with major merger), we simply assume a single burst started at $T_0$, with a length of $\tau$. The schematic of this star formation model is shown in Fig.~\ref{fig:sed}.

We use the \citet{bruzual03} stellar population model, with the redshift fixed to the one estimated with the SDSS DR6, $z=1.07$ \citep[][]{hewett10}, and the \citet{calzetti00} dust attenuation model. It is generally challenging to determine metallicity with only broad band photometry, and thus we fix it to solar. This is reasonable given that galaxies in this mass range are already metal-enriched, as seen at similar or earlier epochs \citep[e.g.,][]{thomas10,onodera15,morishita19,kriek19}.

In Fig.~\ref{fig:sed}, we show examples of the SED results in three different regions; the outer region ($\sim0.9$\,arcsec from the LWC; Voronoi ID 4), the inner ($\sim0.2$\,arcsec from the LWC; Voronoi ID 13), and the blue blob (see Sec.~\ref{ssec:bb}). These examples demonstrate that three data points are sufficient to constrain {\it relative} differences in age of different regions, by capturing both the Balmer break by $V_{\rm 606}-J_{\rm 110}$ and the rest-optical slope by $J_{\rm 110}-H_{\rm 160}$. The best-fit parameters for all tessellated regions are summarized in Table.~\ref{tab:sed}.

\begin{figure}
\centering
	\includegraphics[width=0.46\textwidth]{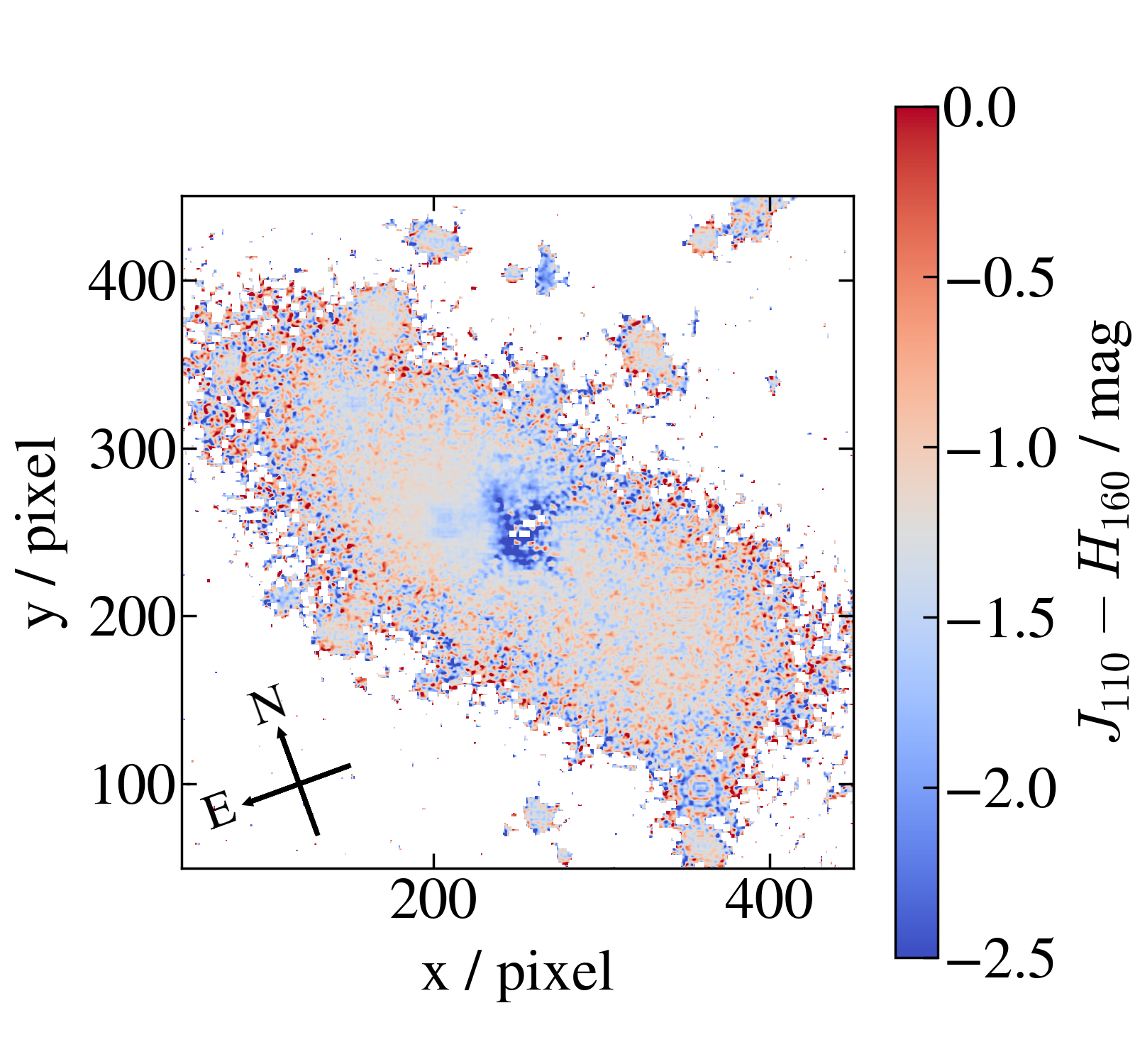}
	\caption{
	Two-dimensional $J_{\rm 110}-H_{\rm 160}$ color map of the host of \ccc. Note that the quasar component is subtracted from each of the images used here. The image exhibits smooth morphology across the host, except two {\it blue} regions near the quasar (mostly residual) and the blue blob (Sec.~\ref{ssec:bb}). The outer regions of the host are excluded and set to blank.
	}
\label{fig:dust}
\end{figure}

A caveat is that with the number of data points used, degeneracy between age and dust is still not completely resolved, which makes inference on absolute values challenging. Our SED fitting results indicate small to moderate attenuation ($A_V\simlt2.0$\,mag). However, in the case of the \ccc\ host we do not expect significant dust attenuation, because no dust lane-like feature is seen in our \hst\ images --- such feature would be clearly seen in the presence of significant dust attenuation \citep[see, e.g.,][for offset AGN candidates with clear dust lanes]{skipper18,hogg21}. In Fig.~\ref{fig:dust}, we show the two-dimensional distribution of $J_{\rm 110}-H_{\rm 160}$ color of the host. While the color does not necessarily represent the exact dust attenuation due to the degeneracy, from the smooth morphology across the host it is unlikely to expect significant dust structures. 

In addition, \citet{podigachoski15} reported tentative ($<3\sigma$) or non-detection of \ccc\ in FIR bands, at 160, 250, 350, and 500\,$\mu$m by the Herschel Space Observatory, implying no significant amount of dust is present in the system. Fixing dust attenuation to zero would increase the age estimates, to account for red components; therefore, our age estimates, especially those with non-zero $A_V$, should be taken as lower limits. 

\begin{figure}
\centering
	\includegraphics[width=0.232\textwidth]{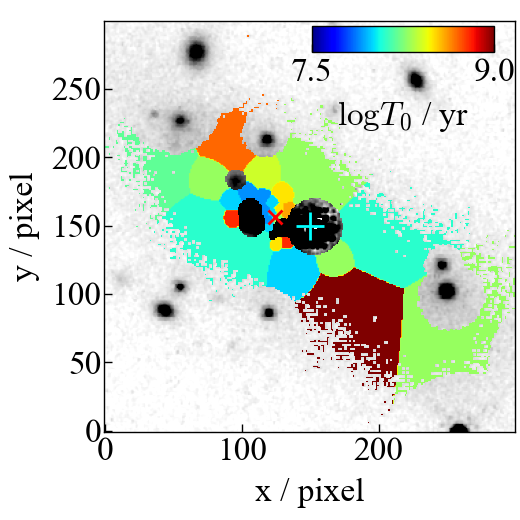}
	\includegraphics[width=0.232\textwidth]{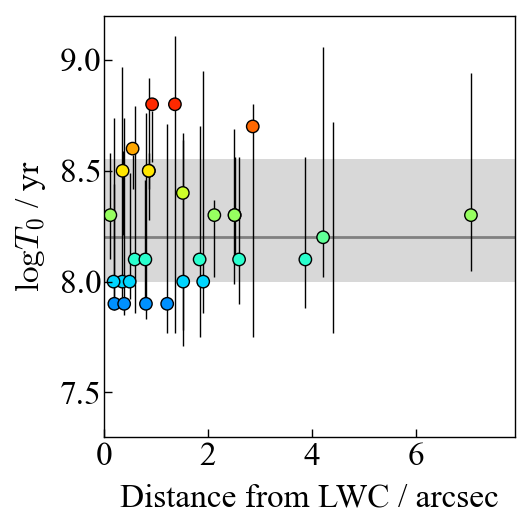}
	\includegraphics[width=0.232\textwidth]{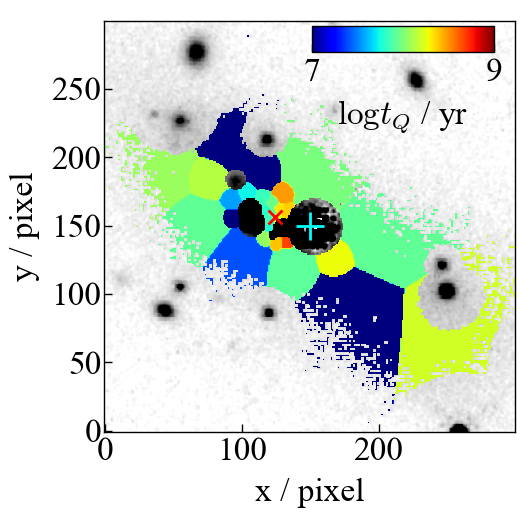}
	\includegraphics[width=0.232\textwidth]{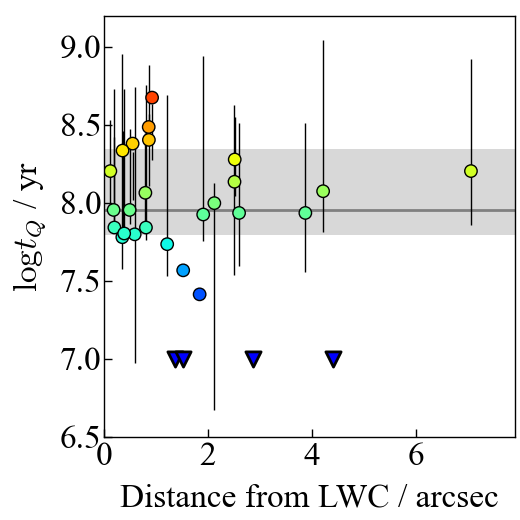}
	\caption{
	Spatial and radial distributions of two parameters, $T_0$ ({\it Top rows}) and $t_Q$ ({\it bottom}) are shown. Colors in the radial distribution plots correspond to those in the spatial distribution plots. Four of the tessellated pixels show on-going star formation (i.e. $t_Q=0$) and are shown at an arbitrary value (blue triangle).
	}
\label{fig:age}
\end{figure}

In Fig.~\ref{fig:age}, we show the spatial distribution of each of the derived parameters. Our interest here is $T_0$, the time at which primary star formation started, and $t_Q$, which we introduce to refer to duration since star formation was truncated ($\equiv T_0-\tau$). There is no clear radial trend for $T_0$, with a median value of $160_{-40}^{+200}$\,Myr, with the associating errors capturing the 16\,/\,84\,th percentiles. Similarly, no clear spatial correlation nor radial trend is observed for $t_Q$. The parameter ranges from $\sim30$\,Myr to $\sim500$\,Myr, with a median value of $t_Q=90_{-30}^{+110}$\,Myr, except a few sub-regions with on-going star formation (i.e. $t_Q \leq 0$). The total mass, derived by summing the best-fit estimate over the Voronoi segments, is found to be $\logm=11.4$.

\subsection{Compact blue star-forming region in the host galaxy}\label{ssec:bb}

As we highlight in the color-composite image of Fig.~\ref{fig:mosaic}, there is a compact star-forming blob (SFB1 hereafter) located at $\sim2$\,arcsec from the quasar to the East-North direction. This blob was seen in the early data and reported in \citet{chiaberge17}, but its properties remained unclear. Our SED modeling revealed its age, $T_0\sim160$\,Myr, and stellar mass, $\sim8\times10^{9}\,M_\odot$ (Fig.~\ref{fig:sed}). We estimate its star formation rate, $65\pm20\,M_\odot {\rm yr^{-1}}$, by averaging over the last 100\,Myr of the best-fit star formation history, which is consistent with the upper limit of the whole system derived from non-detection in Herschel data, $<80\,M_\odot \mathrm{yr^{-1}}$ \citep[][]{podigachoski15}.

The best-fit SFR converts to star formation rate surface density of $\sim1.2\,M_\odot {\rm yr^{-1} kpc^{-2}}$. Considering its stellar surface density ($\log \Sigma_* \sim 8.1\,M_\odot {\rm kpc^{-2}}$), this value is $\sim0.8$\,dex higher than the average value of resolved sub-regions in star-forming galaxies at $z\sim1$ \citep[e.g.,][]{wuyts13}, indicating intense star formation activity may still be ongoing in the host despite relatively old stellar populations elsewhere. Such intense star formation is actually seen in a post-merged system \citep[e.g.,][]{barrows18} but lasts only for a short time scale as $\sim100$\,Myr. We revisit this in Sec.~\ref{sec:dis}. 

Since the template library used in the SED fitting in Section~\ref{ssec:sed} does not include emission lines, we check the fidelity by running another SED fitting code, {\ttfamily gsf} \citep{morishita19}, with emission lines included. While an exponential declining star formation history is used this time, due to the limitation of the code, it derives a consistent value for star formation rate, $70\pm10\,M_\odot {\rm yr^{-1}}$.

While the IFU observations presented in \citet{chiaberge18} covered SFB1, no line was detected, as the observations were originally designed for the brightness of the quasar and exposure was not deep enough to detect any emission lines in SFB1. Recent observations at NOEMA detected extended CO emission near the flux peak position of SFB1 at the same redshift as the optical narrow lines \citep{castignani22}. 

\section{Discussion}\label{sec:dis}

\subsection{On the dynamics of the GW recoil}\label{ssec:dyn}
As we introduced in Sec~\ref{sec:intro}, based on its spectroscopic and imaging properties, \ccc\ was proposed as a clear example of a GW recoiling black hole candidate by \citet{chiaberge17}. With our deep HST observations, we both confirmed and more accurately determined the presence of a spatial offset between the quasar and the host photocenter. Based on our results (Sec.~\ref{ssec:lwc}), we see that the direction of the putative GW kick lies perpendicularly with respect to the radio jet axis. In Fig.~\ref{fig:radio}, we show the 8.44\,GHz radio continuum emission overlaid onto the central region of the host galaxy. The direction of the kick is at the position angle of $\sim+33^{\circ}$, while the radio jet axis has a position angle of $\sim-57^{\circ}$.
As pointed out by, e.g., \citet[][]{lousto12}, the maximum velocity of GW kick ($>2000$\,km/s) may be more likely achieved if the direction of the kick is aligned with the angular momentum of the orbital plane of the merging black hole binary. Since the radio jet axis likely indicates the orientation of the spin of the merged black hole, these results may impose further constraints on the properties of the progenitor binary, in particular with respect to the spin amplitude and relative orientation. A detailed modeling of the binary black hole system and the GW recoil kick lies beyond the scope of this work.

\begin{figure*}
\centering
	\includegraphics[width=0.9\textwidth]{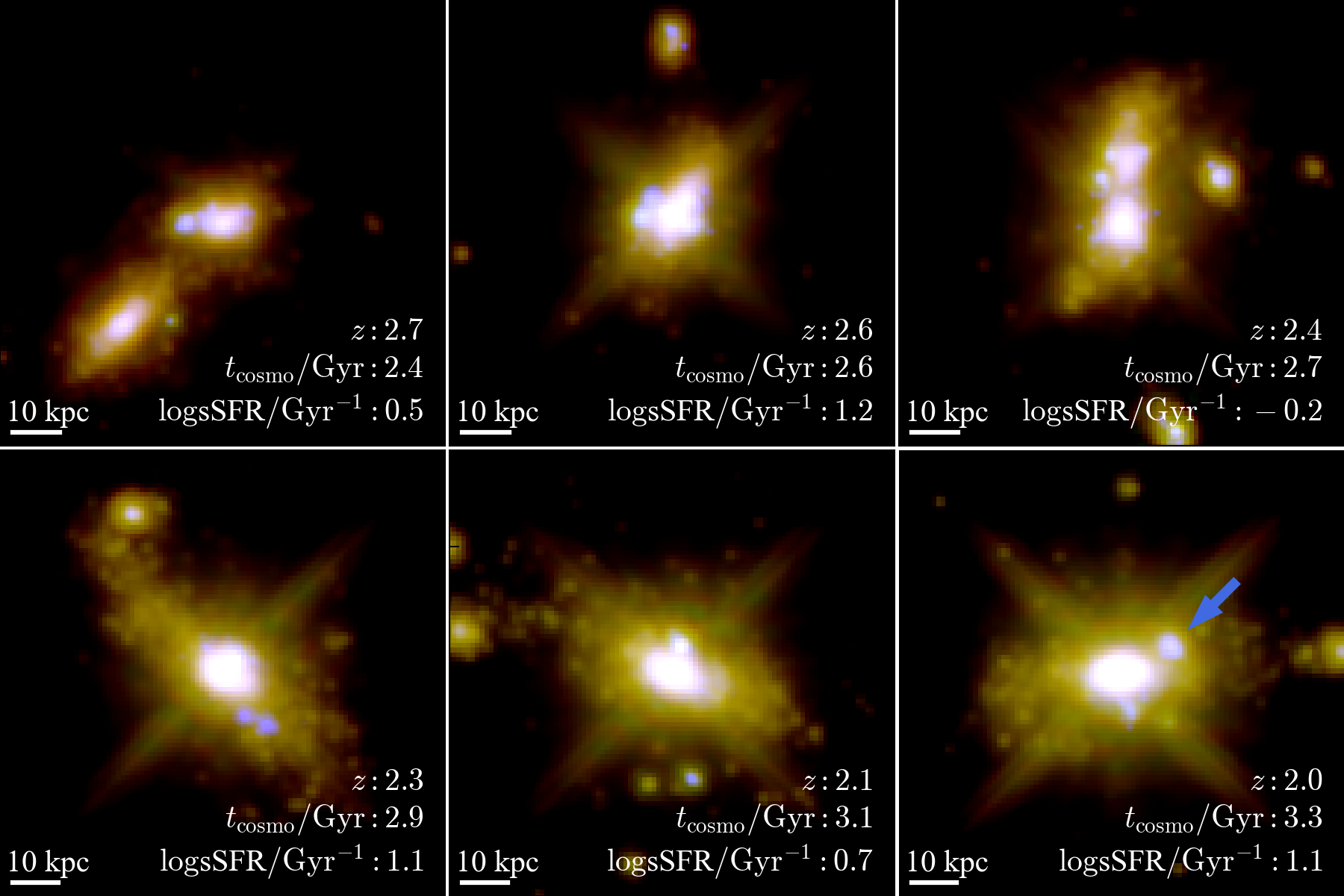}
	\caption{
	A sequence of a merging system VELA04 (Camera05) at $2.0<z<2.7$, taken from the VELA-Sunrise simulation suite. Redshift, cosmic time, and specific star formation rate of each frame are shown in the inset. The RGB composite images consist of F814W, F125W, and F160W, to compensate its higher redshift than \ccc's. 
	One of the star-forming regions (see Sec.~\ref{sec:dis}) is indicated by a blue arrow in the last frame.
	}
\label{fig:merger}
\end{figure*}

\subsection{Insight into the Current Evolutionary Stage }\label{ssec:sims}
In this subsection, we discuss the main results of this work, which focuses on the observed properties of the host. This allows us to speculate the current evolutionary stage of this system.
\citet[][]{chiaberge17} reported a possible shell or tidal feature in the host light profile and speculated its origin as a major merger that occurred about $\sim1$\,Gyr ago from comparison with numerical simulations. Our F160W image presented here revealed even fainter stellar components out to $\sim60$\,kpc from the LWC of the host. While a slight light concentration is seen at the outermost part in the south-east direction, we did not see clear evidence of disturbed morphology or young star clusters in its outskirts, which would be characteristic of on-going merger \citep[e.g.,][]{mulia15}, implying that the system is a relatively old, post-merged remnant. Furthermore, \citet{podigachoski15} derived high SFR in other 3C galaxies at similar redshifts, most of which are associated with on-going mergers, whereas an upper limit was estimated for 3C186 from non-detection in their Herschel data ($<80\,M_\odot \mathrm{yr}^{-1}$). This suggests that \ccc\ is relatively older than other 3C systems in terms of recent star formation too.

To further investigate the current evolutionary phase of \ccc, we compare the observed properties with hydro-dynamical simulations. In Fig.~\ref{fig:merger}, we show a time sequence of a merging system from the VELA-Sunrise project \citep{simons19}. The VELA-Sunrise project provides a set of mock images of simulated galaxies that were originally taken from the VELA simulation suite \citep{ceverino14,zolotov15}. {The simulation suite consists of a variety of galaxies at different evolutionary stages, including both isolated and merging systems. While the original simulations do not include specific recipes for black hole mergers, the datasets offer detailed morphological information of, e.g., a galaxy during merger sequences, at the resolution of HST. Here we select VELA04, one of the merging systems available in the VELA suite. Specifically, the dataset of VELA04 captures the timeline of pre- to post-merger and allows us to investigate the evolutionary stage of \ccc\ by comparing their morphological properties.}

The example shown in Fig.~\ref{fig:merger} is a sequence that captures the pre-merger to post-merged phases of two galaxies at $z\sim2$. The sequence exhibits several features that resemble the observed properties of the \ccc\ host. The last frame of Fig.~\ref{fig:merger} captures elongated (but not disturbed) morphology of the host galaxy and compact star-forming regions in the main body of the host. We also see the presence of young, compact blue regions across the entire sequence. Those blue regions do not sustain more than one simulation frame, implying that their life time is less than its time resolution, $\simlt200$\,Myr. In the last frame of the sequence in Fig.~\ref{fig:merger}, we highlight one such compact blue region. Its specific star formation rate is $\sim40$\,Gyr$^{-1}$ (cf. $\sim8$\,Gyr$^{-1}$ of the blue blob in \ccc). The rest of the stellar populations of the galaxy in the frame is relatively old, with the median age of $\sim1$\,Gyr (cf. $\simgt200$\,Myr of \ccc). {While we also notice morphological differences between the two systems (e.g., more clumps and less relaxed)}, those physical properties are broadly consistent with the observed properties of \ccc\ (Sec.~\ref{ssec:bb}).

The spatial offset found in this study is encouraging regarding a typical timescale after a recoiling event. By using the derived spatial offset of the quasar from the LWC (Sec~\ref{ssec:lwc}) and the velocity offset derived in \citet[][]{chiaberge17}, we estimate that the recoil event occurred $\simgt5$\,Myr ago, though the exact number depends on the assumed kick angle. Theoretically, it is possible for a recoiling BH to remain active for such amount of time. By using the formula in \citet{loeb07} with assumed radiative efficiency of $\epsilon=0.1$ as in  \citet[][]{chiaberge17}, the lifetime of an accretion disk for the black hole mass of \ccc\ is estimated to be $\sim100$\,Myr. In fact, \citet{murgia99} derived the age of the radio source in \ccc, $\sim0.1$\,Myr, from its synchrotron spectrum, which implies that the radio AGN turned on after the recoiling event.

In addition, \citet{blecha11} presented several examples of {simulated} recoiling black holes and estimated a typical lifetime of $\sim10$\,Myr \citep[see also][]{blecha16}. We note, however, that uncertainties in absolute timescale in those simulations are not negligible. Such uncertainties primarily originate in the resolution element, which is typically much larger than the scale of black hole merger (several kpc), making the comparison with observations challenging. This is still the case with more recent simulations that record the exact time of the black hole merger \citep[e.g., TNG;][]{hani20}. Sub-grid models of black hole binary evolution \citep[e.g.,][]{kelley17} show a wide range of black hole merger timescales, from $\sim1$\,Gyr to more than a Hubble time, depending on the model assumptions and the specific conditions in the galaxy nuclei.

\subsection{Alternative Scenario}\label{ssec:alt}
As we mentioned in Sec.~\ref{sec:intro}, there is an alternative scenario for the observed spatial offset, where the quasar is hosted by another, fainter galaxy (``galaxy~B" for convenience) that is superposed on a galaxy at different redshift (``galaxy~A" i.e. the one we have analysed in this study as the host). However, this scenario is unlikely because; (1) we did not confirm any stellar light concentration near the position of the quasar in any filters and (2) there is evidence of narrow absorption at the same redshift of the narrow emission lines. 

For (1), assuming the scaling relation between $M_{*}$ and $M_{\rm BH}$ \citep[e.g.,][]{kormendy13}, stellar mass $M_{*}\sim10^{12}\,M_\odot$ is expected for the black hole mass of \ccc, $M_{\rm BH}\sim3$-$6\times10^9M_\odot$. Such an amount of mass should be noticeably seen in our F160W image regardless of its light profile. \citet{chiaberge17} tested this with various structural parameters for unseen galaxy~B by using {\tt galfit} \citep{pengGALFIT} but concluded that it is not likely not to see such an amount of mass even in their F140W image. The recoiling black hole scenario, which assumes galaxy~A ($\logm\sim11.4$; Sec.~\ref{ssec:sed}) as the host, offers more reasonable agreement in terms of the scaling relation for the stellar and black hole masses observed.
{To further investigate this scenario, we repeat the structural analyses by using our deep F160W image. We provide the original image with the quasar component unsubtracted as the input image to {\tt galfit}. We fit the light profile of the \ccc\ system with one PSF component at the position of the quasar, one S\'ersic component fixed at the LWC position derived above, and one additional S\'ersic component (i.e. galaxy~B) at the quasar position. We fix the position angle and axis ratio of the host to 36 degree and 0.37 \citep{chiaberge17}, respectively, and leave all other parameters as free. We find that the second S\'ersic component does not improve the fit. The best-fit effective radius of this additional component is unphysically small, $54$\,mas. In addition, the total flux of this component remains to be $\sim7\%$ of the host. This converts to stellar mass of $\sim1.8\times10^{10}\,\Msun$ by adopting the same mass-to-light ratio of the host, which is significantly under massive for the black hole mass of \ccc. We therefore conclude that this additional component, if not an artifact such as flux residuals of the PSF component, is unlikely to be the unseen host galaxy of the quasar.}

For (2), \citet{chiaberge17} found narrow absorption features both in \ly\ and \civ\ lines at the same redshift as the narrow emission lines (\oii\ and \neiii). This indicates the presence of gas associated within galaxy~A and thus secures its redshift \citep[see also][who identified CO emission at the same redshift of the narrow lines within the host]{castignani22}. Determining the systemic redshift of galaxy~A directly from stellar absorption lines will be a critical step to test the scenario.

\subsection{Future Prospects}\label{ssec:spec}
Lastly, we note that the observed blueshift measured in the permitted broad lines should carefully be interpreted, as these have large uncertainties primarily due to broad absorption components and blending with other lines \citep[][]{chiaberge17}. For example, such broad absorption is often associated with extreme outflow and seen in a number of local AGN \citep{kaastra14,ebrero16}, making it challenging to attribute the observed blueshift solely to a recoiling event. An ideal line is \hb, which is a good tracer of the kinematics of the accretion disk/inner broad line region {\it and} is isolated from other emission line; thus it can provide a clean measurement of the velocity shift of the accretion disk \citep{chiaberge18}. However, this line is located at $\sim 1\,\mu$m for the redshift (plus possible large blueshift) of \ccc\ and the access to the entire line profile with the current spectrographs may be challenging due to atmospheric absorption and the gap between CCD and IR-detector. The IFU mode of JWST's NIRSpec would be an ideal choice, allowing us to confirm the velocity shift and to locate spatial distribution simultaneously.

High-resolution sub-mm/mm observations of dense gas in \ccc\ would also be a critical step to confirm the physical origin of the blueshift. This is because dense gas in the molecular torus is considered to be orbiting at large distances and thus not gravitationally bounded to the recoiling black hole itself. Indeed, \citet{decarli14} tested this on a candidate, QSO J0927+2943 at $z=0.697$, and identified molecular gas traced by the CO(2-1) line at the redshift of the quasar's broad line, arguing the need for another scenario to explain the properties of the quasar \citep[see also][]{heckman09,shields09}. 

\section{Summary}\label{sec:sum}
In this study, we investigated the properties of the host of \ccc, a recoiling black hole candidate at $z=1.0685$. By carefully analyzing newly taken deep images by HST, we confirmed the previously reported spatial offset, $11.1\pm{0.1}$\,kpc in projected distance, between the core of \ccc\ and the light-weighted center of the host galaxy. We did not find evidence of a recent merger, such as a young starburst in disturbed outskirts. We analyzed the underlining stellar populations of the host and then compared the observed properties with numerical simulations, finding that the observed properties of \ccc\ are consistent with an old merger remnant. Based on those pieces of evidence, we concluded that the recoiling black hole scenario is still a plausible explanation for the puzzling nature of \ccc.

Lastly, we wish to stress that \ccc\ is among the highest redshift candidates of a recoiling black hole that have been identified as of today. Given the higher merger rate at high redshift \citep[e.g.,][]{lotz11,snyder17}, it is reasonable to expect that more candidates will be identified in systematic high-resolution imaging surveys of radio quasars at similar redshifts. Future observations with high-resolution imaging cameras of JWST, even with a short amount of exposure time available in, e.g., Survey programs\footnote{\url{https://jwst-docs.stsci.edu/jwst-opportunities-and-policies/jwst-call-for-proposals-for-cycle-1/jwst-cycle-1-proposal-categories/general-observer-go-proposals/survey-go-proposals}}, will allow us to identify more recoiling black hole candidates and investigate the ubiquity of this extreme phenomenon.

\acknowledgements
We thank the anonymous referee for the constructive comments. We thank Emanuele Berti and Davide Gerosa for fruitful discussion on the physical interpretation of our findings, and Markos Georganopoulos, Andrea Marinucci, Eileen Meyer, Eric S. Perlman, William B. Sparks, Grant R. Tremblay for their contribution to the HST observing program GO-15254. MC thanks Marta Volonteri for providing constructive comments on our study. Support for this study was provided by NASA through a grant HST-GO-15254.005-A from the Space Telescope Science Institute, which is operated by the Association of Universities for Research in Astronomy, Inc., under NASA contract NAS 5-26555. This work is based in part on observations taken by the VLA, operated by the U.S. National Radio Astronomy Observatory, which is a facility of the National Science Foundation, operated under cooperative agreement by Associated Universities, Inc. LB acknowledges support from NSF award AST-1909933. SB and CO acknowledge support from the Natural Sciences and Engineering Research Council (NSERC) of Canada. GC acknowledges the support from the grant ASI n.2018-23-HH.0. 

\tabletypesize{\small}
\tabcolsep=3pt
\startlongtable
\begin{deluxetable}{cccccccc}
\renewcommand{\arraystretch}{.9}
    \tablecolumns{8}
    \tablewidth{0pt}
    \tablecaption{
    Positions of tessellated segments and the best-fit SED parameters.
    }
\tablehead{
    \colhead{Voronoi ID} &
    \colhead{SNR} &
    \colhead{$r^{\dagger}$} &
    \colhead{$\log M_*$} &
    \colhead{$\log T_0$} &
    \colhead{$\log \tau$} &
    \colhead{$\log t_Q$} &
    \colhead{$A_V$}\\
    \colhead{} &
    \colhead{} &
    \colhead{arcsec} &
    \colhead{$\log M_\odot$} &
    \colhead{yr} &
    \colhead{yr} &
    \colhead{yr} &
    \colhead{mag}
    }
\startdata
0 & 19.0 & 0.59 & 9.2 & 8.1 & 7.8 & 7.8 & 1.7 \\
1 & 22.8 & 0.80 & 9.6 & 8.1 & 7.0 & 8.1 & 1.5 \\
2 & 26.7 & 1.91 & 10.0 & 8.0 & 7.2 & 7.9 & 1.2 \\
3 & 23.1 & 0.86 & 9.6 & 8.5 & 7.8 & 8.4 & 1.2 \\
4 & 23.9 & 4.40 & 10.7 & 9.4 & 9.8 & ---$^\ast$ & 1.3 \\
5 & 21.0 & 0.92 & 9.4 & 8.8 & 8.2 & 8.7 & 0.0 \\
6 & 23.1 & 2.51 & 9.9 & 8.3 & 7.0 & 8.3 & 1.1 \\
7 & 23.4 & 0.35 & 9.2 & 8.0 & 7.6 & 7.8 & 1.3 \\
8 & 22.3 & 7.06 & 10.6 & 8.3 & 7.6 & 8.2 & 2.1 \\
9 & 18.7 & 0.20 & 9.0 & 7.9 & 7.0 & 7.8 & 1.0 \\
10 & 19.3 & 0.12 & 9.2 & 8.3 & 7.6 & 8.2 & 0.9 \\
11 & 22.5 & 1.84 & 10.0 & 8.1 & 8.0 & 7.4 & 1.7 \\
12 & 23.4 & 1.36 & 9.7 & 8.8 & 9.4 & ---$^\ast$ & 1.6 \\
13 & 15.7 & 0.18 & 8.9 & 8.0 & 7.0 & 8.0 & 0.9 \\
14 & 20.3 & 0.55 & 9.3 & 8.6 & 8.2 & 8.4 & 0.0 \\
15 & 18.2 & 0.39 & 9.0 & 7.9 & 7.2 & 7.8 & 1.0 \\
16 & 16.4 & 0.36 & 9.1 & 8.5 & 8.0 & 8.3 & 0.4 \\
17 & 22.1 & 2.60 & 10.3 & 8.1 & 7.6 & 7.9 & 2.1 \\
18 & 17.9 & 0.49 & 9.2 & 8.0 & 7.0 & 8.0 & 1.2 \\
19 & 21.5 & 0.81 & 9.5 & 7.9 & 7.0 & 7.8 & 1.6 \\
20 & 26.0 & 1.52 & 9.7 & 8.0 & 7.8 & 7.6 & 1.8 \\
21 & 20.8 & 1.21 & 9.6 & 7.9 & 7.4 & 7.7 & 1.8 \\
22 & 26.1 & 3.87 & 10.4 & 8.1 & 7.6 & 7.9 & 1.6 \\
23 & 22.3 & 2.51 & 10.2 & 8.3 & 7.8 & 8.1 & 1.9 \\
24 & 22.4 & 0.86 & 9.5 & 8.5 & 7.0 & 8.5 & 0.0 \\
25 & 23.8 & 1.52 & 9.8 & 8.4 & 9.6 & ---$^\ast$ & 1.7 \\
26 & 18.1 & 2.86 & 10.1 & 8.7 & 10.0 & ---$^\ast$ & 1.8 \\
27 & 15.4 & 4.21 & 10.3 & 8.2 & 7.6 & 8.1 & 2.2 \\
28 & 24.9 & 2.12 & 10.1 & 8.3 & 8.0 & 8.0 & 0.6 \\
\enddata
\tablecomments{
    $\dagger$: Distance from the light-weighted center.\\
    $\ast$: Regions with on-going star formation (i.e. $t_Q = 0$).
    }
\label{tab:sed}
\end{deluxetable}

\noindent
{\it Software:} Astropy \citep{astropy:2013, astropy:2018}, Astrodrizzle \citep{hack12}, FAST \citep{kriek09}, GALFIT \citep{peng02, peng10}, gsf \citep{morishita19}, LACOSMIC \citep{vandokkum01}, numpy \citep{oliphant2006guide,van2011numpy}, Pypher \citep{boucaud16}, SExtractor \citep{bertin96}.


\clearpage

\bibliographystyle{apj}
\bibliography{./adssample}

\end{document}